\documentclass[12pt]{iopart}

\usepackage{graphicx}

\begin{document}

\title{High-beta equilibrium in mirror machine with population of fast sloshing ions}

\author{Ivan S. Chernoshtanov}

\address{Budker Institute of Nuclear Physics SB RAS, 11, akademika Lavrentieva prospect, Novosibirsk, 630090
Russia} \ead{I.S.Chernoshtanov@inp.nsk.su} \vspace{10pt}
\begin{indented}
\item[]December 1 2025
\end{indented}

\begin{abstract}

A method of constructing the high-$\beta$ (diamagnetic-bubble-like)
equilibrium with a population of fast sloshing ions is discussed.
Fast ions move along betatron orbits; such ions can arise because of
off-axis neutral beam injection. Conservation of the adiabatic
invariant $\oint p_rdr$ of these ions is proposed; a simplified
expression for the invariant is presented. Numerical examples of
equilibrium with sloshing ions are shown and conservation of the
invariant is justified by a direct numerical simulation of motion of
fast ions even in the case with $\beta\approx1$.

\end{abstract}

%
\noindent{\it Keywords}: mirror machine, diamagnetic confinement,
plasma equilibrium, particle dynamics in electromagnetic fields
%
%
%
%

\section{Introduction}

Regimes with an extremely high plasma pressure, which is equal to
the pressure of the external magnetic field, are of interest.
Besides achieving the maximal thermonuclear power \cite{Budker58},
this regime makes it possible to minimize the cyclotron radiation
and suppress essentially the longitudinal losses
\cite{Beklemishev16}. Potential advantages of the regimes with a
high-beta plasma were stimulated numerous efforts to attain such
regimes experimentally. It necessary to note the Astron experiment
\cite{Briggs75}, experiments with cusps (see \cite{Haines77} and
\cite{Ioffe84}), and experiments with mirror machines with
high-power neutral beam injection (NBI). The first attempt to
achieve this regime with $\beta\approx1$ in a mirror machine with
NBI was made in the 2XIIB device in 1979 \cite{Turner89}; it was
shown that the plasma pressure was limited because of the anomalous
diffusion of the fast ions driven by the ion-cyclotron oscillation
and non-axisymmetric geometry of the magnetic field. An experiment
with powerful normal NBI in an axisymmetric mirror is now planned in
the Compact Axisymmetic Toroid (CAT) device
\cite{Bagryansky16,Davydenko18}. First successful demonstrations of
a formation of configuration with $\beta\approx1$ in a mirror
machine with skew NBI were performed recently in the Norm device
\cite{Roche2025}.

Accumulation of fast ions arising because of NBI should result in
the formation of a Field Reversed Configuration (FRC)
\cite{Steinhauer11} or so-called diamagnetic bubble configuration
\cite{Beklemishev16}. Regions with a weak magnetic field arise in
both cases; it may result in non-conservation of the magnetic
momentum and chaotic behavior of the fast ions. In an axisymmetric
mirror machine, the azimuthal momentum $p_\theta$ is conserved and
an ion on chaotic trajectory can be confined if the azimuthal
component of velocity is large enough and the direction of rotation
of the ion coincides with the direction of the cyclotron rotation
(so-called co-orbit motion \cite{Morozov98}); it is absolute
confinement \cite{Morozov98,VTP63}. Moreover, arising of a weak
magnetic field region may result in chaotization of the dynamics of
the target ions and restrict the lifetime of the target plasma. Of
course, chaotization of the dynamics of ions has a negative effect,
which restricts the plasma parameters. Rough estimations of the
particle and energy lifetimes in axisymmetric configurations with
$\beta\approx1$ are given in
\cite{Haines77,Hsiao85,Chernoshtanov2022}. Typically, the
assumptions that most ions move chaotically in regimes with
$\beta\approx1$ is used for theoretical description of equilibrium
and transport; see, for example \cite{Steinhauer11Pt,Khristo25}.

The motion of an ion in this high-beta mirror machine is accompanied
by radial oscillations of the ion (due to the reflection from
regions with a strong magnetic field, as well as usual cyclotron
rotation). When the frequency of these radial oscillations is large
enough, the adiabatic invariant $\oint p_rdr$ can be conserved (in a
low-beta machine this invariant transforms to the usual magnetic
momentum) \cite{Firsov58,Tsidulko16}. Of course, conservation of the
invariant is a desirable effect because it can restrict the losses
of fast ions. So, the invariant conservation in high-beta
magnetically confined plasmas is studied in several theoretical
works (see, for example, review \cite{Haines77}, where the positive
effect of adiabatic invariant conservation is demonstrated, and
article \cite{Egedal18}, where conservation of the invariant in a
given magnetic field (the Soloviev solution \cite{Soloviev76}) is
studied in detail).

It is obvious that at $\beta\approx0$ almost all ions move
regularly. So, ions with regular behavior should be present during
the transition from the mirror machine with a moderate $\beta$ to
the regime with $\beta\approx1$. Moreover, it is noted in
\cite{Chernoshtanov2022} that ions moving along betatron orbits
(those are close to a circle; the ions rotate in the direction which
coincides with the cyclotron rotation, for example, see figure 1b in
\cite{Rostoker02}) have a high frequency of radial oscillations
(which is close to the ion cyclotron frequency in a vacuum magnetic
field), and thus these ions tend to move regularly. The adiabatic
invariant for these ions is destroyed later than for other ions. The
ions on betatron orbits can arise because of off-axis NBI (as in the
described experiments with an NBI-sustained high-beta plasma), and
one can expect that the population of fast ions in these experiments
can be confined in the adiabatic regime up to $\beta\approx1$.

In this article we present a method of calculation of axisymmetric
equilibriums in a high-beta plasma with a population of
adiabatically moving fast ions. These equilibriums can be useful for
reconstruction of distribution of plasma pressure and density in
experiments, as well as for analysis of magnetohydrodynamic and
kinetic instabilities in high-beta plasmas. The article is organized
as follows. In the second part, the expression for the adiabatic
invariant for ions moving along a betatron orbit is discussed, and
the current generated by the population of these ions is calculated.
In the third part, some methods of solving numerically the
Grad-Shafranov equation (based on paper \cite{Rostoker02d}) are
described. Some numerical examples are given in the fourth part. The
results of a numerical simulation of the dynamics of fast ions,
which demonstrate that the ions really move regularly in these
axisymmetric high-beta configurations, are also shown in the fourth
part.

\section{Distribution of sloshing ions}

The adiabatic invariant can be written in the following form:
\begin{eqnarray*}
I_r(\varepsilon,p_\theta,p_z)=\oint
p_rdr=2\int_{r_{\min}}^{r_{\max}}dr\sqrt{2m\varepsilon-\frac{(p_\theta-e\Psi(r,z)/c)^2}{r^2}-p_z^2},
\end{eqnarray*}
where $e$ and $m$ is the ion charge and mass, respectively;
$\Psi(r,z)=\int_0^rr'B_z(r',z)dr'$ is the magnetic flux;
$p_\theta=mrv_\theta+(e/c)\Psi$ is the azimuthal momentum (which is
a constant of motion in the case of an axisymmentic magnetic field);
$p_r$ and $p_z$ is the radial and longitudinal components of the ion
momentum, respectively; $r_{\min}$ and $r_{\max}$ are the roots of
the integrand.

We will consider ions with a small radial component of velocity
later. These ions move along betatron orbits with $r\approx
r_\beta$. At a given $p_\theta$ and $z$, the function
$(p_\theta-(e/c)\Psi)^2/r^2$ has a minimum at $r=r_\beta$ (namely,
$p_\theta-(e/c)\Psi=-mr_\beta^2\Omega(r_\beta)$). The frequency of
small radial oscillations of the ions is the betatron frequency
$\Omega_\beta=(\Omega\partial_r(r\Omega))^{1/2}$, where
$\Omega(r,z)=eB(r,z)/(mc)$ is the local cyclotron frequency of the
ions \cite{Wong91}. Assuming $r\approx r_\beta$, one can expand the
integrand near $r_\beta$:
\begin{eqnarray*}
I_r\approx2m\int_{r_{\min}}^{r_{\max}}dr\sqrt{\frac{2\varepsilon}{m}-\frac{p_z^2}{m^2}-r_\beta^2\Omega^2(r_\beta)-\Omega_\beta^2(r-r_\beta)^2}=\frac{\pi
m}{\Omega_\beta}\left(\frac{2\varepsilon}{m}-r_\beta^2\Omega^2(r_\beta)-\frac{p_z^2}{m^2}\right).
\end{eqnarray*}
So, in this approximation, the adiabatic invariant is proportional
to the ratio of the square of the radial velocity to the betatron
frequency.

It is convenient to modify the expression for the invariant. Namely,
let $r_{\beta\min}(p_\theta)$ be the radius of the betatron orbit at
the longitudinal coordinate $z=z_{\min}$ where the magnitude of the
vacuum magnetic field is minimal. Besides,
$\Omega_{\min}=\Omega(r_{\beta\min},z_{\min})$ and
$\Omega_{\beta\min}=\Omega_\beta(r_{\beta\min},z_{\min})$. We define
a new invariant $V_z$ by the relation
\begin{eqnarray*}
I_r=\frac{\pi
m}{\Omega_\beta}\left(\frac{2\varepsilon}{m}-r_\beta^2\Omega^2(r_\beta)-\frac{p_z^2}{m^2}\right)\equiv\frac{\pi
m}{\Omega_{\beta\min}}\left(\frac{2\varepsilon}{m}-r_{\beta\min}^2\Omega_{\min}^2-V_z^2\right)
\end{eqnarray*}
so that
\begin{eqnarray*}
V_z^2=\left(\frac{2\varepsilon}{m}-r_{\beta\min}^2\Omega_{\min}^2\right)-\frac{\Omega_{\beta\min}}{\Omega_\beta}\left(\frac{2\varepsilon}{m}-r_{\beta}^2\Omega^2-\frac{p_z^2}{m^2}\right).
\end{eqnarray*}
The invariant $V_z$ is the longitudinal component of the ion
velocity at $z=z_{\min}$.

We choose the distribution of the fast ions as the sum of the
following model distributions:
\begin{equation}
F_n(\varepsilon,p_\theta,V_z)=\delta\left(\frac{2\varepsilon}{m}-v_0^2\right)\frac{p_0\delta(p_\theta+p_0)}{\Gamma(n+1/2)w_\|}\left(\frac{V_z^2}{w_\|^2}\right)^ne^{-V_z^2/w_\|^2},\label{Fn00}
\end{equation}
where $v_0>0$, $p_0>0$, and $w_\|>0$ are parameters, $n\geq0$ is an
integer, and $\Gamma(x)$ is the gamma function. The parameter $p_0$
fixes the radius of the betatron orbit of ions $r_{\beta\min}$; the
distribution (\ref{Fn00}) has a peak at $V_z=w_\|\sqrt{n}$. So, the
radial component of the ion velocity is of the order of
$(v_0^2-\Omega_{\min}^2r_{\beta\min}^2-w_\|^2n)^{1/2}$ and can be
made small through a proper choice of the parameter $v_0$. The sum
of functions (\ref{Fn00}) allows us to approximate any sufficiently
smooth function of $V_z$.

The concentration of the ions with distribution (\ref{Fn00}) is
\begin{eqnarray*}
n_n(r,z)=\frac{(-1)^np_0H(v_0^2-V_\theta^2(r,z))}{\Gamma(n+1/2)mrw_\|}\left.\frac{\partial^n}{\partial\alpha^n}\left(
e^{-(v_0^2-r_{\beta\min}^2\Omega_{\min}^2-(\Omega_{\beta\min}/\Omega_\beta)(v_0^2-r_{\beta\min}^2\Omega_{\min}^2))\alpha/w_\|^2}\right.\right.\nonumber\\
\left.\left.e^{-(\Omega_{\beta\min}/\Omega_\beta)(v_0^2-V_\theta^2(r,z))\alpha/(2w_\|^2)}
I_0\left(\alpha\frac{\Omega_{\beta\min}}{\Omega_\beta}\frac{v_0^2-V_\theta^2(r,z)}{2w_\|^2}\right)\right)\right|_{\alpha=1},
\end{eqnarray*}
where $V_\theta(r,z)=-(p_0+(e/c)\Psi(r,z))/r$ is the azimuthal
component of the velocity of ion with $p_\theta=-p_0$ at the point
with the coordinates $r$ and $z$; $H(x)$ is the Heaviside step
function and $I_0(x)$ is the modified Bessel function. The azimuthal
component of the current density of these ions is
\begin{equation}
j_n(r,z)=eV_\theta(r,z)n_n(r,z).
\end{equation}

\subsection{Spatial distribution of fast ions in vacuum field}

To illustrate the properties of distribution (\ref{Fn00}) let us
consider a low-beta plasma. In this case,
$\Omega_\beta\approx\Omega$, the fast ions move along the field
lines: $r_\beta\approx r_{\beta\min}\sqrt{\Omega_{\min}/\Omega}$.
The longitudinal component of velocity $v_\|$  depends on $z$ as
$v_z^2=V_\|^2-(\Omega/\Omega_{\min}-1)r_{\min}^2\Omega_{\min}^2$ for
an ion with the azimuthal momentum
$p_\theta=-mr_{\min}^2\Omega_{\min}+(e/c)\Psi(r_{\min},z_{\min})$
(the azimuthal component of velocity at $z=z_{\min}$ is
$-r_{\min}\Omega_{\min}$), the energy
$\varepsilon=m(V_\|^2+r_{\min}^2\Omega_{\min})/2$ and the invariant
$V_\|$. The spatial distribution of the density of ions with similar
$p_\theta$, $\varepsilon$, and $V_\|$ can be found from the
condition of conservation of particle flow $nv_\|r_\beta\delta
r=\mbox{const}$ (here $\delta r\sim\Omega^{1/2}$ is the radial width
of the distribution):
\begin{equation}
n(r,z)\approx
n_0\frac{V_\|r_{\beta\min}\delta(r-r_\beta)}{\sqrt{V_\|^2+(1-\Omega/\Omega_{\min})r_{\beta\min}^2\Omega_{\min}^2}}.\label{Bet00}
\end{equation}
The density has a singularity at the turning points, where
$\Omega/\Omega_{\min}=(2\varepsilon/m)/(r_{\min}^2\Omega_{\min}^2)$
and $v_z=0$.

In section 4 we consider fast ions with a similar $p_\theta$ and a
distribution independent on $V_\|$ at $0\leq|V_\||\leq V_{\max}$. To
find the spatial distribution of the density and azimuthal current
of these ions in this vacuum field, we integrate distribution
(\ref{Bet00}) over $V_\|$:
\begin{eqnarray*}
n(r,z)\approx
n_0r_{\beta\min}\delta(r-r_\beta)\sqrt{V_{\max}^2+\left(1-\frac{\Omega}{\Omega_{\min}}\right)r_{\beta\min}^2\Omega_{\min}^2},\quad
j_\theta(r,z)\approx-er_\beta\Omega n(r,z).
\end{eqnarray*}
The density monotonically decreases from $z_{\min}$ to the mirrors
and the azimuthal current has a maximum at $z\neq z_{\min}$ if
$V_{\max}>r_{\beta\min}\Omega_{\min}$.

\section{Numerical scheme}

Let's consider an equilibrium of plasma with a population of fast
sloshing ions. The plasma is surrounded by a cylindrical vacuum
chamber with radius $r_w$; the distance between the magnetic mirrors
is $L$; the vacuum magnetic field has the minimum $B_{\min}$ at
$z=z_{\min}$. We neglect the currents and electric fields generated
by the target plasma. To find an equilibrium one should solve the
Grad-Shafranov equation
\begin{equation}
\Delta^*\Psi=r\frac{\partial}{\partial
r}\frac1r\frac{\partial\Psi}{\partial
r}+\frac{\partial^2\Psi}{\partial z^2}=\frac{4\pi}{c}rj_\theta.
\label{G00}
\end{equation}
To solve numerically this equation we use a modification of the
method described in \cite{Rostoker02d}. Namely, the equation can be
rewritten in the form of an integral equation:
\begin{eqnarray*}
\Psi_p(r,z)=\frac{4\pi}{c}\int
G(r,r',z-z')j_\theta(r',\Psi_p(r',z')+\Psi_v(r',z'))dr'dz',\label{G01}
\end{eqnarray*}
where $\Psi_p$ is the flux generated by the fast ions, $\Psi_v$ is
the flux of the vacuum magnetic field (and thus $\Delta^*\Psi_v=0$),
and $G(r,r',z)$ is Green's function for equation (\ref{G00}), which
satisfies the equation $\Delta^*G=r'\delta(r-r')\delta(z)$.
Physically, Green's function is the flux generated by a thin
circular coil with radius $r'$. We neglect the current flowing over
the vacuum chamber surrounding the plasma and assume that the system
is periodical in $z$ with a period $L$. In this case, Green's
function can be expressed through complete elliptic integrals of the
first and the second kinds $K(\xi)$ and $E(\xi)$:
\begin{eqnarray*}
G(r,r',z)=\sum_n\sqrt{\frac{4rr'}{\xi_n}}\frac{2E(\xi_n)+(\xi_n-2)K(\xi_n)}{4\pi},\quad\xi_n=\frac{4rr'}{(r+r')^2+(z+nL)^2}.
\end{eqnarray*}

Let's introduce a rectangular grid along $r$ and $z$; $r_i=\Delta
r(i-1)$ and $z_j=-L+\Delta z(j-1)$ for $1\leq i\leq n_r$ and $1\leq
j\leq n_z$, where $\Delta r=r_w/(n_r-1)$ and $\Delta z=2L/(n_z-1)$.
We solve integral equation (\ref{G01}) by iterations:
\begin{eqnarray*}
\Psi_{i,j}^{(n)}=(1-\alpha)\Psi_{i,j}^{(n)}+\alpha\frac{4\pi}{c}\frac{J_\theta}{J^{(n-1)}}\sum_{k=1}^{n_r}\sum_{l=1}^{n_r}G_{i,k,|j-l|+1}j_\theta\left(r_k',\Psi_{k,l}^{(n-1)}+\Psi_v(r_k',z_l')\right)\Delta
r\Delta z,
\end{eqnarray*}
where $\Psi_{i,j}=\Psi_p(r_i,z_j)$, $G_{i,k,j}=G(r_i,r_k',z_j)$,
$J_\theta>0$ is the full azimuthal current of the sloshing ions
through the semi-plane $r,z$,
\begin{eqnarray*}
J^{(n)}=\sum_{k=1}^{n_r}\sum_{l=1}^{n_r}j_\theta\left(r_k',\Psi_{k,l}^{(n)}+\Psi_v(r_k',z_l')\right)\Delta
r\Delta z,
\end{eqnarray*}
the parameter $0<\alpha\leq1$ is used to stabilize numerical
instabilities (typically we choose $\alpha=0.1\div0.2$), and the
zero approximation is $\Psi_{i,j}^{(0)}=\Psi_v(r_i,z_j)$.

Green's function has a singularity at $r=r'$ and $z=z'$. To avoid
this singularity, we integrate Green's function near the node at
$i=k$ and $j=1$ and thus
\begin{eqnarray*}
G_{i,i,1}=-\log\left(\frac{\Delta r^2+\Delta
z^2}{(16r_i)^2}\right)-\frac{\Delta z}{\Delta
r}\arctan\left(\frac{\Delta r}{\Delta z}\right)-\frac{\Delta
r}{\Delta z}\arctan\left(\frac{\Delta z}{\Delta r}\right).
\end{eqnarray*}

The impact parameter $r_{\beta\min}$ is chosen equal for all
sloshing ions (and therefore the parameter
$p_0=-mr_{\beta\min}^2\Omega_{\min}$ is equal for the ions). The sum
of distribution functions (\ref{Fn00}) with the same $w_\|$ and
different $n$ and
$v_0=(r_{\beta\min}^2\Omega_{\min}^2+w_\|^2n)^{1/2}$ is used for
modeling the distribution function of the sloshing ions.

\section{Numerical example}

To illustrate the main properties of a high-$\beta$ axisymmetric
equilibrium with sloshing ions, we present the results of a
numerical solution to Grad-Shafranov equation (\ref{G00}). The
magnetic system is a set of coaxial coils with radius of 52.5 cm;
the distance between the coils is $L=200$ cm; the current in each
coil is 700 kA. These coils generate a corrugated magnetic field
with magnitude at the center $B_v=2$ kGs and mirror ratio of 4. The
distribution of the fast ions $f_i$ is the sum of functions
(\ref{Fn00}):
$f_i(\varepsilon,p_\theta,V_\|)=\sum_{n=0}^{n_{\max}}c_nF_n(\varepsilon,p_\theta,V_\|;v_n,p_*,w_\|)$
with $p_*=mr_*^2\Omega_{\min}-(e/c)\Psi(r_*,z_{\min})$ (so,
$r_{\beta\min}=r_*$); $v_n=(r_*^2\Omega_{\min}^2+nw_\|^2)^{1/2}$;
$r_*=10$ cm (so, the plasma radius is equal to 10 cm); and
$w_\|/\Omega_v=2$ cm. The coefficients are as follows: either
$c_{2n}=\Gamma(2n+1/2)/\Gamma(2n+1)$ and $c_{2n+1}=0$ at $n>0$ and
$c_0=\sqrt{\pi}/2$ (in this case the distribution weakly depends on
$V_\|$ if $|V_\||<w_\|\sqrt{n_{\max}}$; we will name this
distribution flat) or $c_{2n}=\Gamma(2n+1/2)2n/\Gamma(2n+1)$ and
$c_{2n+1}=0$ at $n>0$ and $c_0=\sqrt{\pi}/2$ (in this case the
distribution is peaked near $V_\|=w_\|\sqrt{n_{\max}}$; we will name
it a peaked distribution).

\subsection{Structure of magnetic field}

\begin{figure}[!h]
\begin{center}
\includegraphics[width=0.495\textwidth]{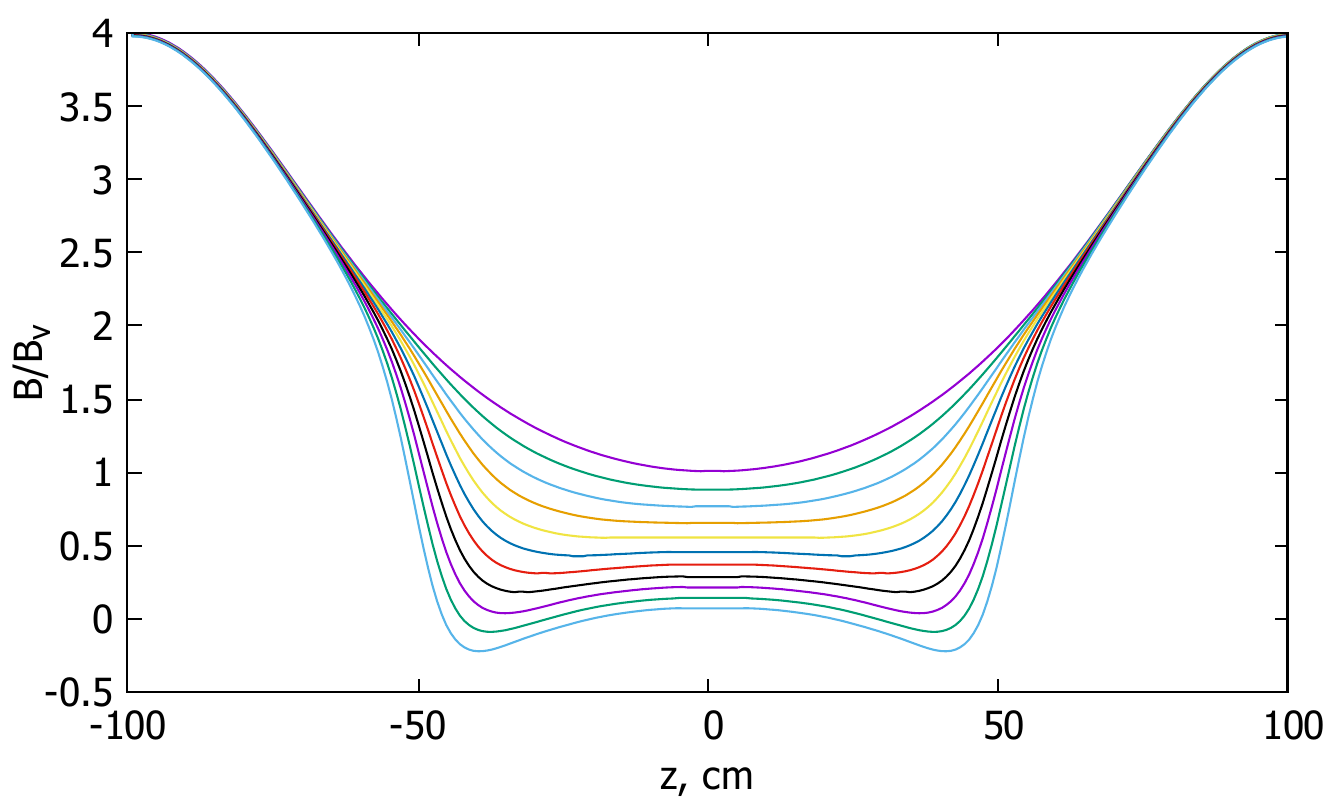}
\includegraphics[width=0.495\textwidth]{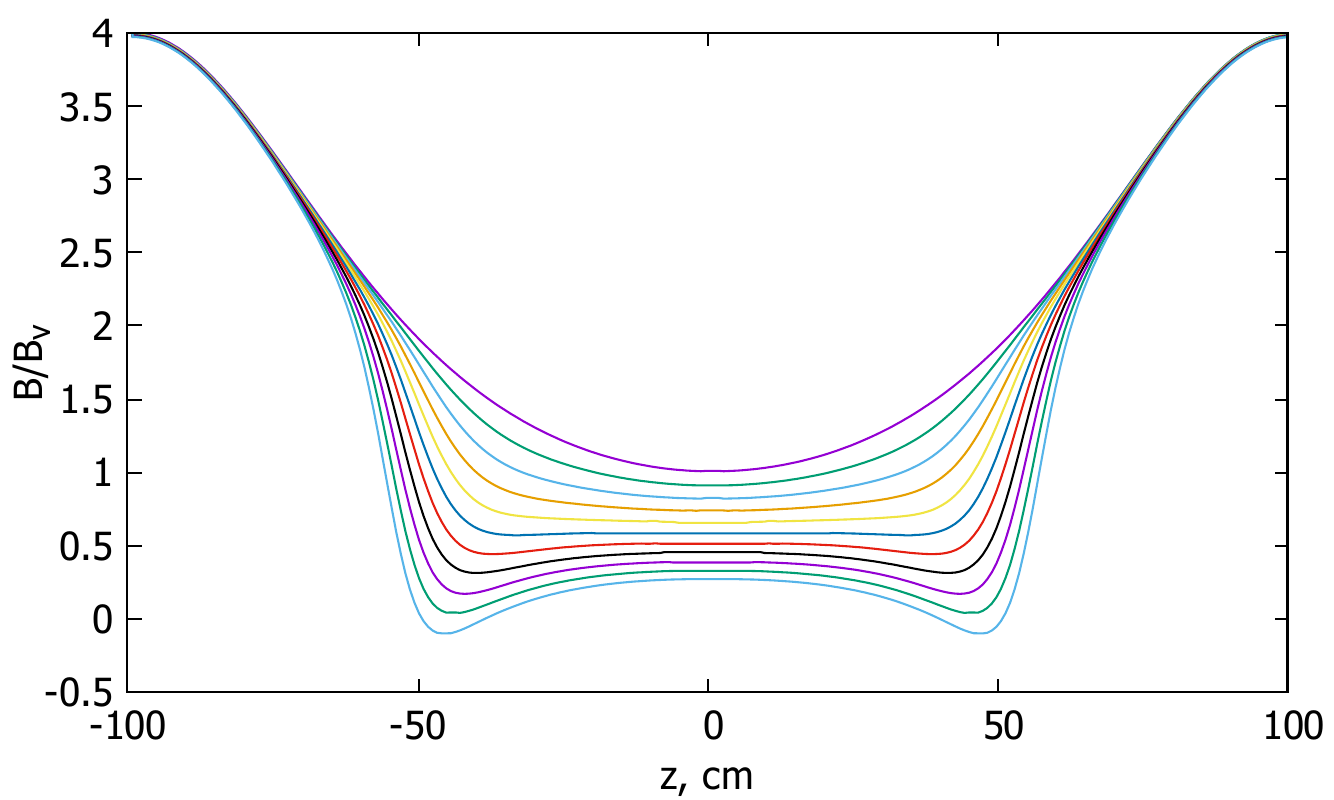}
\end{center}
\caption{\label{Bz} Examples of longitudinal distribution of
magnitude of magnetic field at axis. Flat (left) and peaked (right)
distributions with $n_{\max}=25$. Full azimuthal current $J_\theta$
varies from zero to 270 kA with step of 24.5 kA.}
\end{figure}

The profiles of the magnetic field on the axis at different
$J_\theta$ are shown in figure \ref{Bz}. The sloshing ions spend a
lot of time in the vicinity of the turning points, where
$v_z\approx0$. It results in a concentration of the azimuthal
current generated by the fast ions in the vicinity of the turning
points and smoothing of the profile of the magnetic field between
the turning points at high $\beta$. A peaking of the sloshing ions
can result in field reversing near the turning points, but this
effect seems to be nonphysical because of the destruction of the
adiabatic invariant of fast ions in configurations with the reversed
field (see below).

\begin{figure}[!h]
\begin{center}
\includegraphics[width=0.495\textwidth]{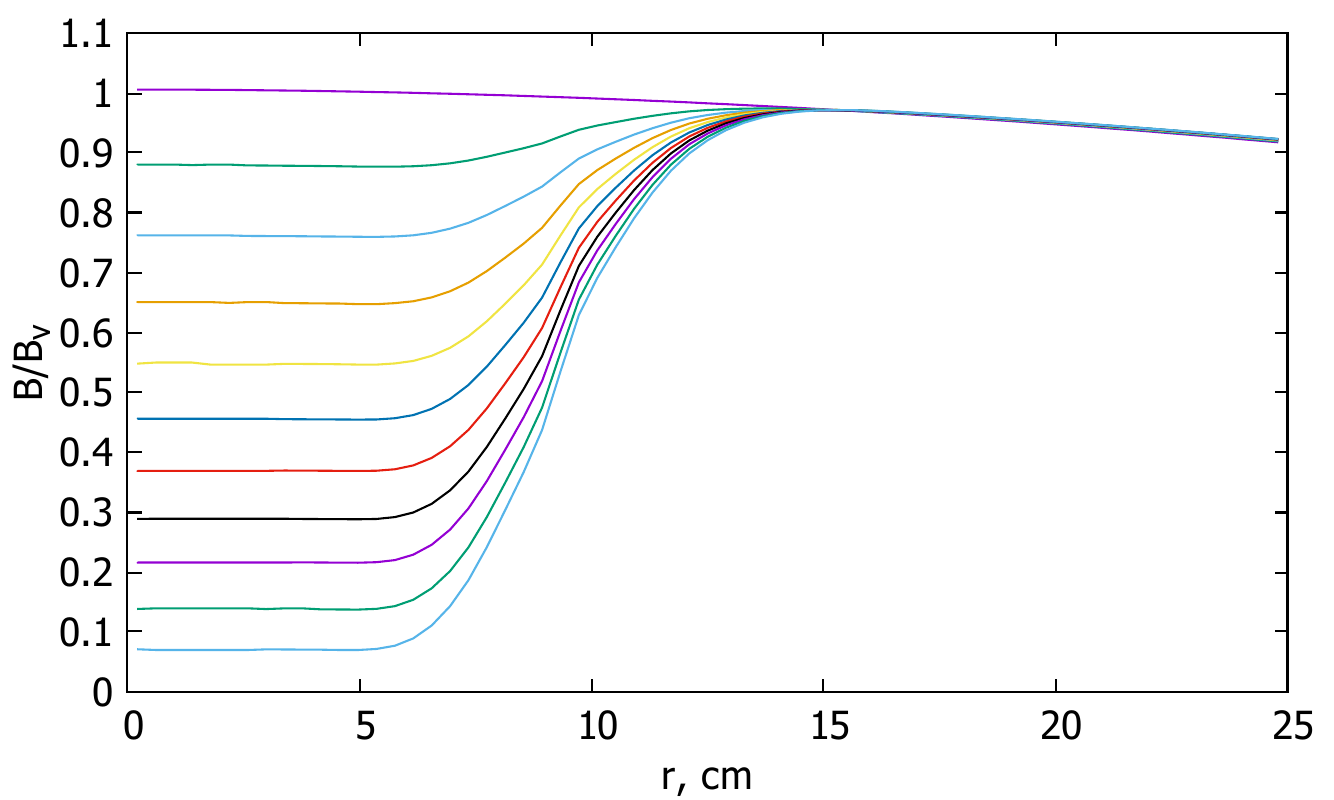}
\includegraphics[width=0.495\textwidth]{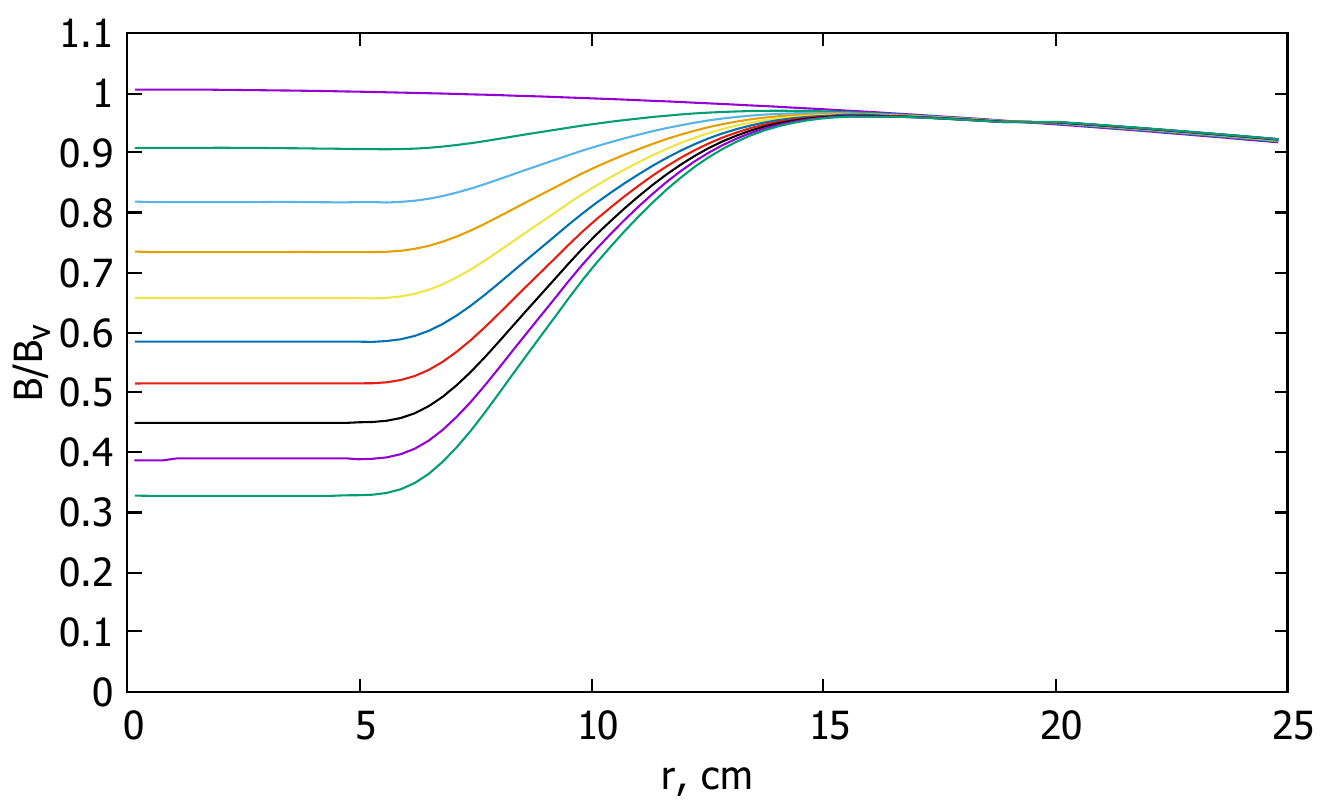}
\end{center}
\caption{\label{BzR} Examples of radial distribution of
$z$-component of magnetic field at $z=0$ for flat (left) and peaked
(right) distributions. Parameters are the same as in figure
\ref{Bz}.}
\end{figure}

\begin{figure}[!h]
\begin{center}
\includegraphics[width=1.0\textwidth]{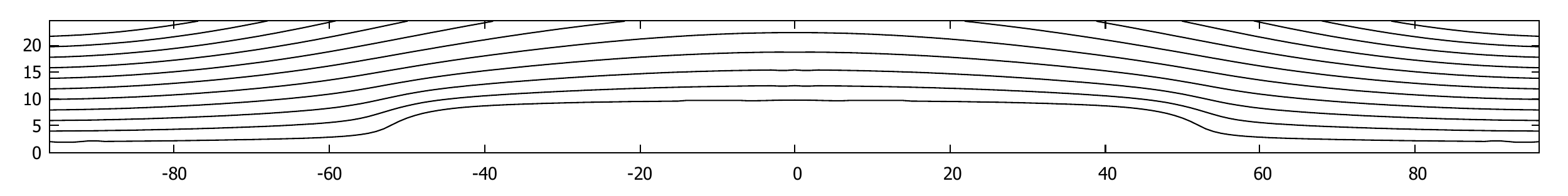}
\end{center}
\caption{\label{FieldLines} Examples of field lines in case of flat
distribution with $n_{\max}=25$ and $J_\theta=270$ kA.}
\end{figure}

An example of magnetic field lines is shown in figure
\ref{FieldLines}. An important difference between the high-$\beta$
configuration with sloshing ions and the FRCs is the absence of
strongly inclined field lines in regions with high and moderate
magnitudes of the magnetic field. The magnetic field generated by
fast ions in vicinity of the turning points is the same as the field
of a long solenoid with radius equal to the plasma radius. So,
behind the turning points, the magnetic field rises from zero to the
magnitude of the vacuum field in a distance approximately equal to
the plasma radius, and the slope of the field lines does not exceed
45$^\circ$. Therefore, the magnetic field varies smoothly, which
allows the adiabatic invariant to be conserved.

\subsection{Uniqueness of equilibrium}

\begin{figure}[!h]
\begin{center}
\includegraphics[width=0.5\textwidth]{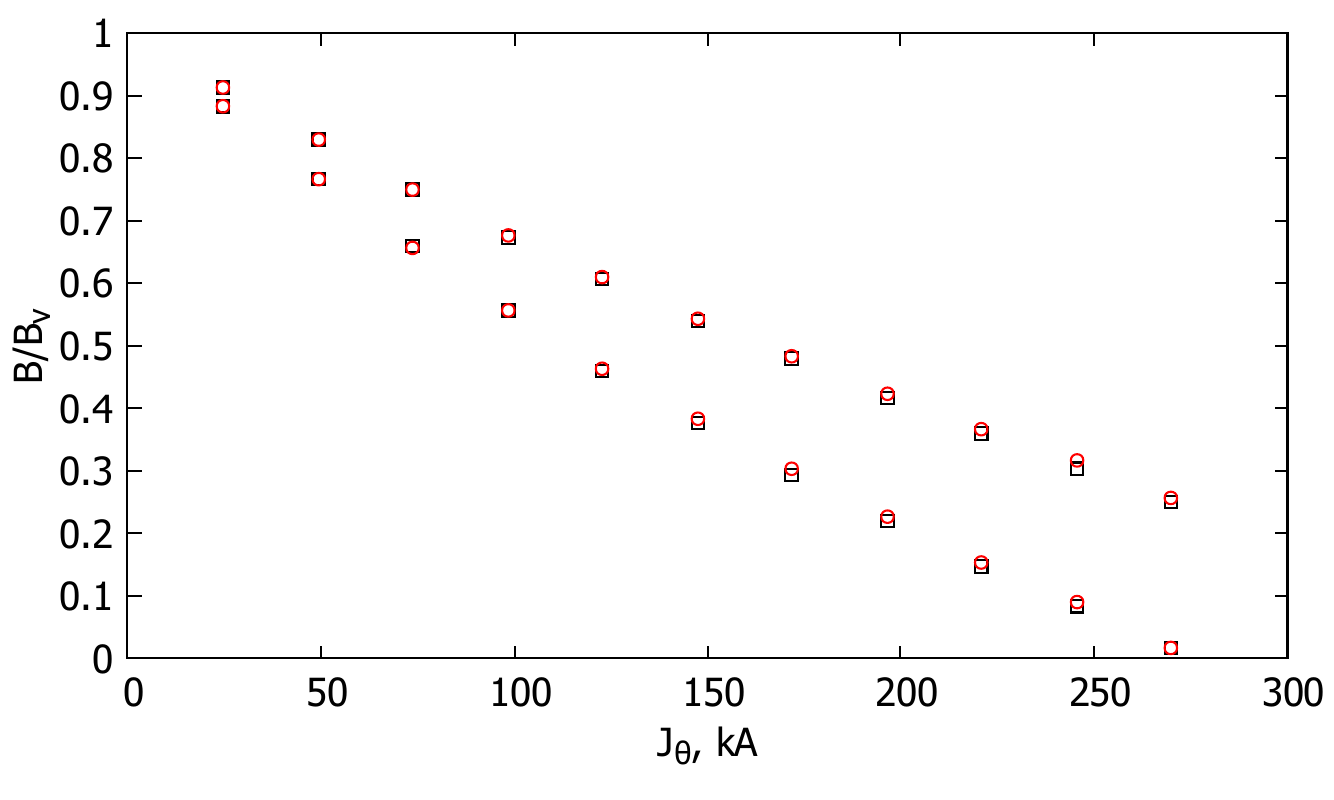}
\end{center}
\caption{\label{BJt} Dependence of magnetic field magnitude in
center in case of flat (upper points) and peaked (lower points)
distributions. Zero approximation: vacuum flux (squares) and
magnetic field for $J_\theta=270$ kA (circles).}
\end{figure}

The Grad-Shafranov equation is non-linear, and thus it can have
several solutions. It means that in principle different equilibriums
can exist at the same distribution functions of fast ions and
parameter $J_\theta$ (it should be noted that transitions between
different high-beta equilibriums are described in
\cite{Kotelnikov10}). To verify this assumption we compare the
results of numerical solutions with different zero approximations
$\Psi_{i,j}^{(0)}$: the vacuum magnetic flux and the magnetic flux
for an equilibrium with the full azimuthal current $J_\theta=1.65$
MA. The dependence of the magnetic field magnitude at $r=0$ and
$z=z_{\min}$ is shown in figure \ref{BJt}; there are no differences
between the solutions. This indicates that the equilibrium is
uniquely determined by the distribution function and full azimuthal
current.

\subsection{Conservation of adiabatic invariant}

The described high-beta equilibriums exist only if the adiabatic
invariant is conserved for most fast ions. This can be expected from
general considerations: in an axisymmetric field the invariant can
be conserved if the frequency of the radial oscillations (the
betatron frequency) is large enough in comparison with $v_\|/l$,
where $v_\|$ and $l$ are the characteristic values of the
longitudinal component of the velocity and the longitudinal scale of
magnetic field, respectively. However, the magnetic field changes
most sharply in the vicinity of the turning points (where $l$ is of
the order of the plasma radius $r_*$), where the longitudinal
velocity of ions $v_\|$ is small. Moreover, ions with a large
betatron radius and a small radial velocity spend a significant
amount of time in the region with a moderate magnitude of the
magnetic field; so, the betatron frequency of these ions (which is
of the order of the cyclotron frequency in the vacuum field) is
large enough in comparison with $v_\|/r_*$. So, the invariant is
expected to be conserved for most of the fast ions, but this should
be verified numerically.

For the verification, a direct numerical simulation of the ion
motion and a construction of Poincare mappings were performed. For
approximation of the magnetic flux and their derivatives, a special
approximation (see Appendix) is used.

\begin{figure}[!h]
\begin{center}
\includegraphics[width=0.5\textwidth]{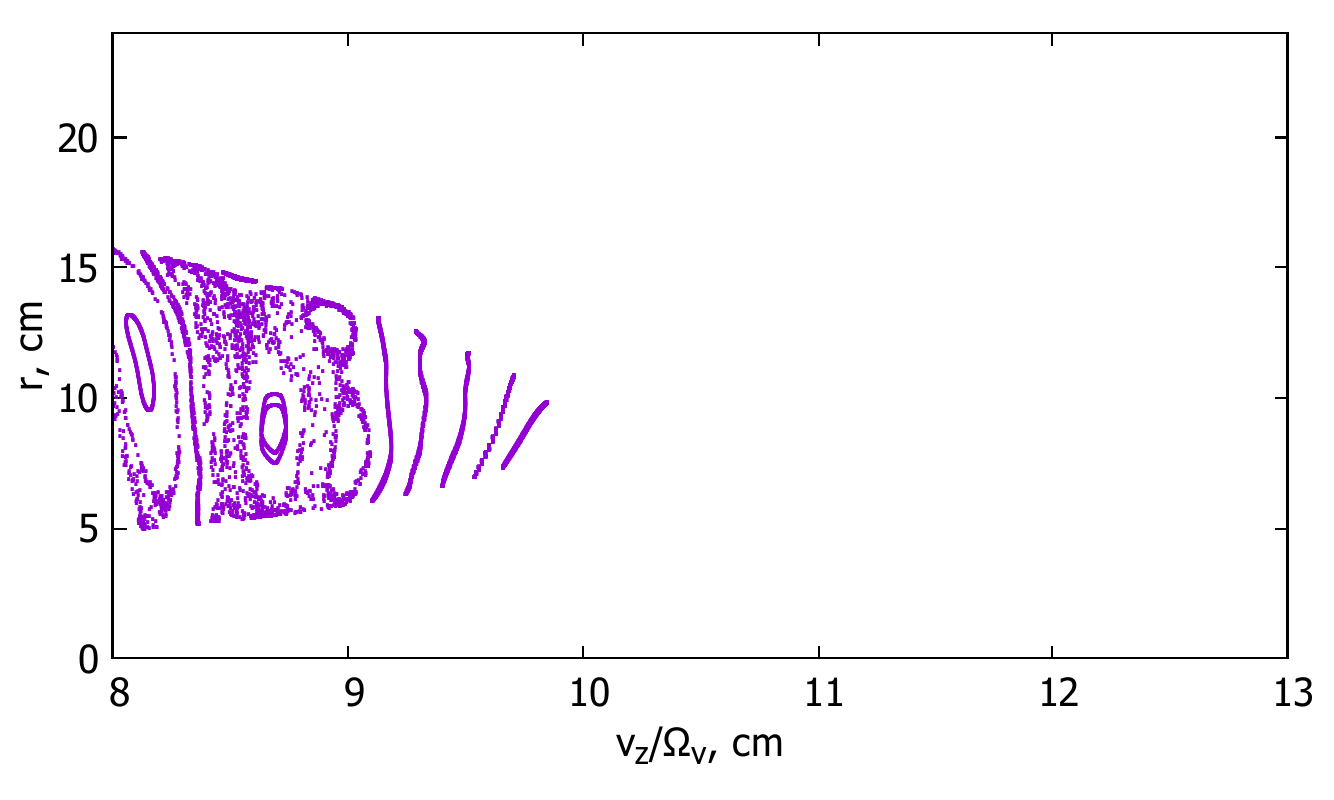}
\includegraphics[width=0.5\textwidth]{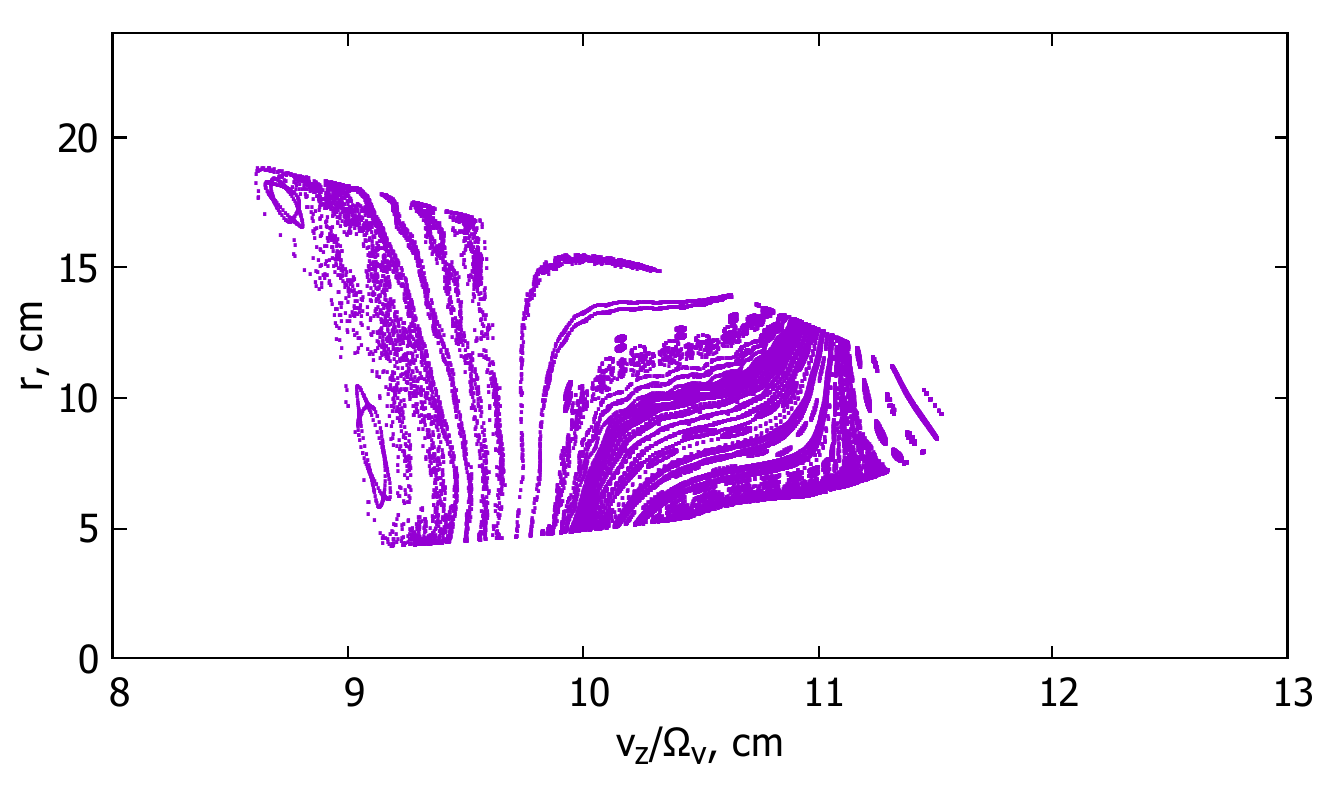}
\includegraphics[width=0.5\textwidth]{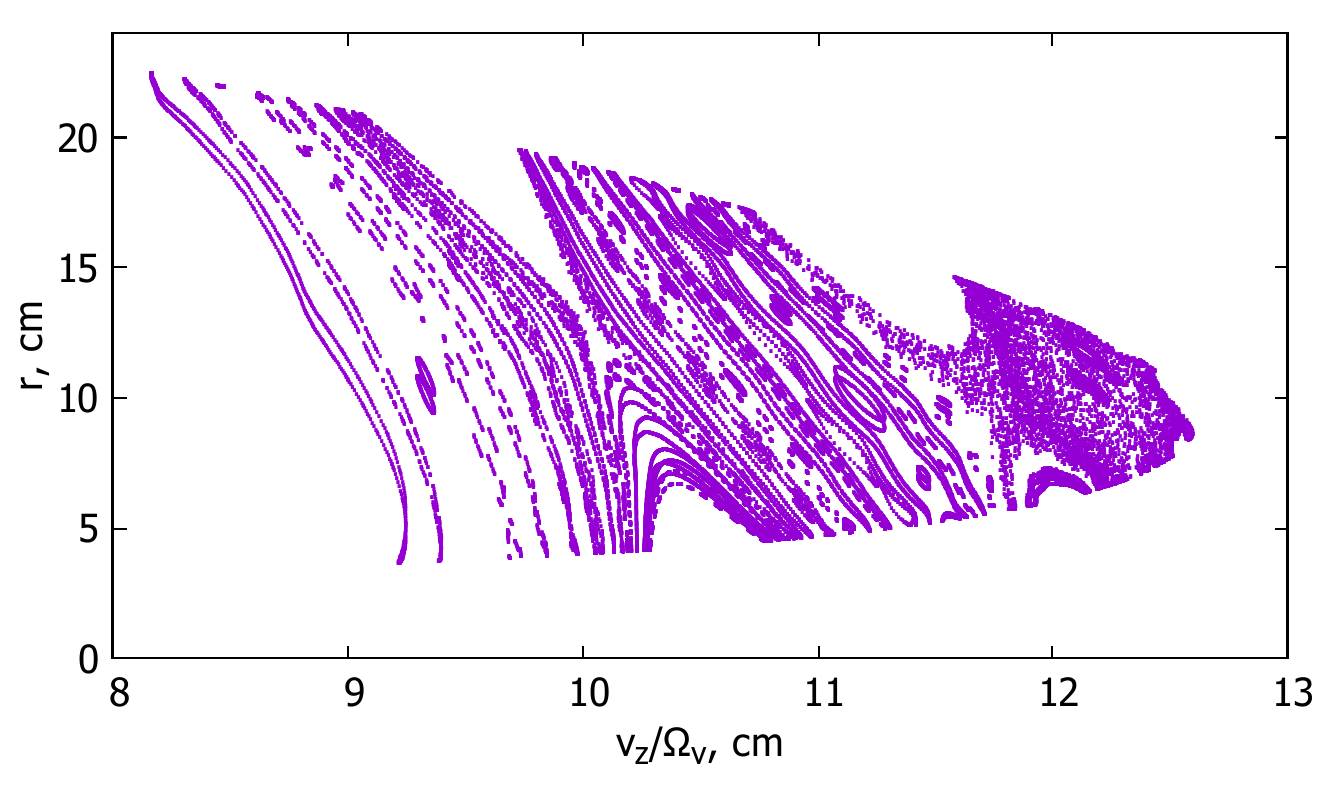}
\end{center}
\caption{\label{Poincare} Examples of Poincare mapping for ions with
azimuthal momentum $p_\theta/(m\Omega_v)=-39$ cm$^2$ and kinetic
energy $\sqrt{2\varepsilon/(m\Omega_v^2)}=11$ cm (upper),
$\sqrt{2\varepsilon/(m\Omega_v^2)}=12.5$ cm (middle) and
$\sqrt{2\varepsilon/(m\Omega_v^2)}=13.5$ (down). Magnetic field are
the same as in figure \ref{FieldLines}.}
\end{figure}

Typical Poincare mappings are shown in figure \ref{Poincare}. The
mappings show the dependence of the longitudinal component of ions
on the radial coordinate at the moments when the ions intersect the
plane $z=z_{n_z/2}$; all ions start from $r=10$ cm with a similar
kinetic energy and azimuthal momentum. If the kinetic energy of the
ion does not exceed $m(r_*^2\Omega_{\min}^2+n_{\max}w_\|^2)/2$ (this
energy is approximately equal to the maximum energy of ions in the
distribution of the sloshing ions) and the longitudinal component of
ion velocity is large enough, then trajectory of the fast ion are
regular (see figure \ref{Poincare}). Only resonances between the
longitudinal and radial motions are observed; these resonances are
surrounded by thin stochastic layers. If the energy of a fast ion is
is too large, then ions with the largest longitudinal velocity move
chaotically. One can conclude that almost all fast ions, which
generate azimuthal current, move regularly even if magnetic field on
the axis is near zero.

\section{Conclusion}

The fast ions arising because of off-axis NBI can be confined in the
adiabatic regime even if $\beta\approx1$ because of the following
features of these orbits: the ions move predominantly in the region
with a moderate magnetic field and the frequency of the radial
oscillations of these ions is relatively high. The conservation of
the invariant can influence the equilibrium (for example, sloshing
ions tend to cluster in the vicinity of the turning points) and thus
the stability and transport of matter and energy in the high-beta
plasma.

A population of sloshing ions can influence the plasma stability
different ways. On the one hand, such population can provoke
excitation of different kinetic instabilities, such as the
axisymmetic Global acoustic mode \cite{Skovorodin13}, analogs of the
Alfv{\'e}n instability with the azimuthal wavenumber $m=-1$ (see,
for example, \cite{Casper82}), etc. (isotropization of the
distribution function of fast ions due to the destruction of the
adiabatic invariant and chaotic motion is in turn to stabilize these
instabilities). On the other hand, concentration of sloshing ions in
the vicinity of turning points makes conducting stabilizers (of
course located near the turning points) more effective. Clarifying
these question requires further consideration.

The numerical solution of the Grad-Shafranov equation and the
simulation of the fast ions collisionless dynamics have been
performed at the computing cluster of the BINP SB RAS (Novosibirsk,
Russia).

\section*{Appendix. Approximation of magnetic flux}

To explicitly take into account the flow behavior in the axis
vicinity we use the following approximation:
\begin{eqnarray*}
\Psi(r,z)=r^2\sum_{j=1}^{n_r}\sum_{k=1}^{n_z}\frac{\Psi_{j,k}}{r_j^2}\left(W_n\left(\frac{r}{\Delta
r}-j+1\right)+\delta_{j,1}W_n\left(\frac{r}{\Delta
r}+1\right)\right)W_n\left(\frac{z-L/2}{\Delta z}-k+1\right),
\end{eqnarray*}
where
\begin{eqnarray*}
W_n(x)=\frac{1}{(n-1)!}\sum_{k=1}^{\lfloor(n+3)/2\rfloor}(-1)^kC_n^kH\left(\frac{n}{2}-k-|x|\right)\left(\frac{n}{2}-k-|x|\right)^{n-1}.
\end{eqnarray*}

\begin{figure}[!h]
\begin{center}
\includegraphics[width=0.5\textwidth]{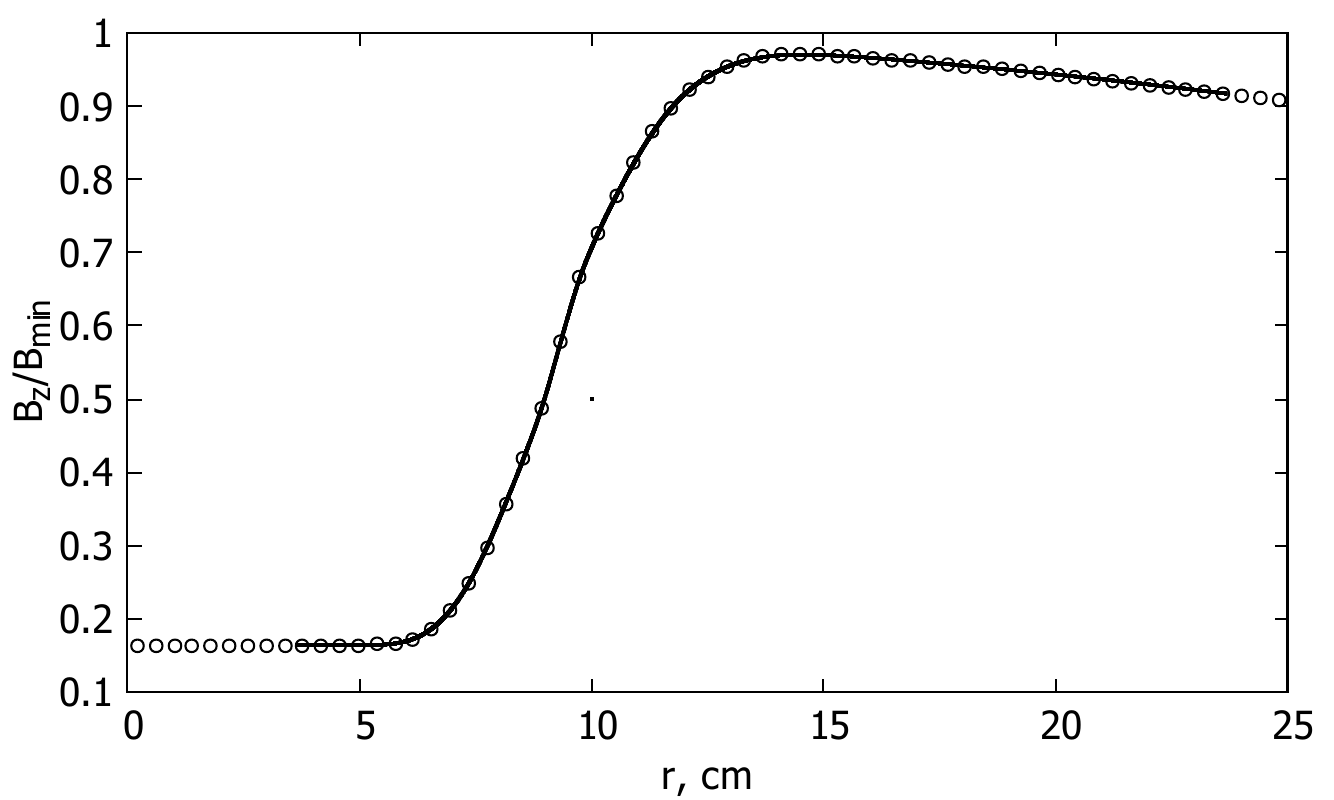}
\end{center}
\caption{\label{BzA} Examples of $z$-component of magnetic field at
points where ions intersect plane $z=z_{n_z/2}$ (points) and
difference derivative of flux at points $(r_{i+1}+r_i)/2$
(circles).}
\end{figure}

We use the three-point approximation with $n=4$. In this case, the
first derivatives of the flux (through which the magnetic field
components are expressed) have continuous derivatives. The accuracy
of the approximation is controlled by comparison of the
$z$-component of the magnetic field at the points where ions
intersect the plane $z=z_{n_z/2}$ and the values of the difference
derivative of the flux
$2(\Psi_{i+1,n_z/2}-\Psi_{i,n_z/2})/((r_{i+1}+r_i)\Delta r)$ at the
points $r=(r_{i+1}+r_i)/2$ (see figure \ref{BzA}).

\par


\begin{thebibliography}{99}

\bibitem{Budker58}
{\it G.I. Budker.} Fizika Plazmy i Problema Upravlyaemykh
Termoyadernykh Reaktsii Vol. 3 (Pod red. M A Leontovicha) (M.:
Izd-vo AN SSSR, 1958) p. 3; Per. na angl. yaz., {G.I. Budker} Plasma
Physics And The Problem Of Controlled Thermonuclear Reactions Vol. 3
(Ed. M. A. Leontovich) (New York: Pergamon Press, 1959) p. 3

\bibitem{Beklemishev16}
{\it A.D. Beklemishev.} Diamagnetic "bubble" equilibria in linear
traps. Physics of Plasmas {\bf 23}, 082506 (2016).
https://doi.org/10.1063/1.4960129

\bibitem{Briggs75}
{\it Briggs, R.J., Hester R. E., Porter G. D., Sherwood W.A.,
Spoerlein R., Stallard B.W., Taska J., Weiss P.B.} ASTRON Program
Final Report, Lawrence Livermore Laboratory Report UCRL-51874
(1975).

\bibitem{Haines77}
{\it M. G. Haines.} Plasma containment in cusp-shaped magnetic
fields. {\bf 17}, 811 (1977).
https://doi.org/10.1088/0029-5515/17/4/015

\bibitem{Ioffe84}
{\it M. S. Ioffe, B. I. Kanaev, V. V. Piterskii, and E. E.
Yushmanov}, Sov. J. Plasma Phys {\bf 10}, 261 (1984).

\bibitem{Turner89}
{\it W. C. Turner, J. F. Clauser, F. H. Coensgen, D. L. Correll, W.
F. Cummins, R. P. Freis, R. K. Goodman, A. L. Hunt, T. B. Kaiser, G.
M. Melin, W. E. Nexsen, T. C. Simonen, and B. W. Stallard.}
Field-reversal experiments in a neutral-beam-injected mirror
machine. Nuclear Fusion, {\bf 19}, 1011 (1989).
https://doi.org/10.1088/0029-5515/19/8/002

\bibitem{Bagryansky16}
{\it P. A. Bagryansky, T. D. Akhmetov, I. S. Chernoshtanov, P. P.
Deichuli, A. A. Ivanov, A. A. Lizunov, V. V. Maximov, V. V.
Mishagin, S. V. Murakhtin, E. I. Pinzhenin, V. V. Pikhodko, A. V.
Sorokin, and V. V. Oreshonok.} Status of the experiment on magnetic
field reversal at BINP. AIP Conference Proceedings. {\bf 1771},
030015 (2016). https://doi.org/10.1063/1.4964171

\bibitem{Davydenko18}
{\it V. I. Davydenko, P. P. Deichuli, A. A. Ivanov and  S. V.
Murakhtin.} Neutral Beam Injection System for the CAT Experiment.
Plasma and Fusion Research: Regular Articles Volume 14, 2402024
(2019). https://doi.org/10.1585/pfr.14.2402024

\bibitem{Roche2025}
{\it T. Roche et. al.} Generation of field-reversed configurations
via neutral beam injection. Nature Communications, {\bf 16}, 3487
(2025)

\bibitem{Steinhauer11}
{\it L. C. Steinhauer.} Review of field-reversed configurations.
Physics of Plasmas, {\bf 18}, 070501 (2011).
https://doi.org/10.1063/1.3613680

\bibitem{Morozov98}
{\it A. I. Morozov and V. V. Savel'ev.} On Galateas - magnetic traps
with plasma-embedded conductors. Phys.-Usp. 41, 1049 (1998).
https://doi.org/10.1070/PU1998v041n11ABEH000501

\bibitem{VTP63}
{\it L.S. Soloviev.} Geometry of magnetic field, in {\it Reviews of
Plasma Physics}, Vol. 2, edited by M.A. Leontovich (Consultants
Bureau, New York, 1968) p. 3, translated from Russian: Voprosy
Teorii Plasmy (Atomizdat, Moscow, 1963)

\bibitem{Hsiao85}
{\it M. Y. Hsiao and G. H. Miley} Velocity-space particle loss in
field-reversed configurations. Physics of Fluids, {\bf 28}, 1440
(1985). https://doi.org/10.1063/1.864978

\bibitem{Chernoshtanov2022}
{\it I. S. Chernoshtanov}. Collisionless Particle Dynamics in
Diamagnetic Trap. Plasma Physics Reports, {\bf 48}, 2, 79-90 (2022).
https://doi.org/10.1134/S1063780X22020052

\bibitem{Steinhauer11Pt}
{\it L. C. Steinhauer.} Hybrid equilibria of field-reversed
configurations. Physics of Plasmas {\bf 18}, 112509 (2011).
https://doi.org/10.1063/1.3660674

\bibitem{Khristo25}
{\it M. S. Khristo and A. D. Beklemishev.} Plasma equilibrium in
diamagnetic trap with neutral beam injection. Journal of Plasma
Physics, Volume 91 , Issue 1 , February 2025 , E3,
https://doi.org/10.1017/S0022377824001417

\bibitem{Egedal18}
{\it J. Egedal et. al.} Theory of ion dynamics and heating by
magnetic pumping in FRC plasma. Physics of Plasmas, {\bf 25}, 072510
(2018). https://doi.org/10.1063/1.5041749

\bibitem{Firsov58}
{\it O.B. Firsov.} Fizika Plazmy i Problema Upravlyaemykh
Termoyadernykh Reaktsii Vol. 3 (Pod red. M A Leontovicha) (M.:
Izd-vo AN SSSR, 1958) p. 3; Per. na angl. yaz., {O.B. Firsov} Plasma
Physics And The Problem Of Controlled Thermonuclear Reactions Vol. 3
(Ed. M. A. Leontovich) (New York: Pergamon Press, 1959) p. 259

\bibitem{Tsidulko16}
{\it Yu. A. Tsidulko.} Adiabatic model of field reversal by fast
ions in an axisymmetric open trap. Plasma Physics Reports, {\bf 42},
559-565 (2016). https://doi.org/10.1134/S1063780X16060088

\bibitem{Soloviev76}
{\it L.S. Soloviev.} Review of Plasma Physics {\bf 6}, 239 (1976)

\bibitem{Rostoker02}
{\it N. Rostoker and A. Qerushi.} Equilibrium of field reversed
configurations with rotation. I. One space dimension and one type of
ion. Phys. Plasmas. {\bf 9}, 3057 (2002).
https://doi.org/10.1063/1.1475683

\bibitem{Rostoker02d}
{\it A. Qerushi and N. Rostoker.}Equilibrium of field reversed
configurations with rotation. III. Two space dimensions and one type
of ion. Phys. Plasmas. {\bf 9}, 5001 (2002);
https://doi.org/10.1063/1.1517294

\bibitem{Wong91}
{\it H. Vernon Wong, H. L. Berk, R. V. Lovelace, and N. Rostoker.}
Stability of annular equilibrium of energetic large orbit ion beam.
Physics of Fluids B: Plasma Physics. {\bf 3}, 2973 (1991).
https://doi.org/10.1063/1.859931

\bibitem{Kotelnikov10}
{\it I. A. Kotelnikov, P. A. Bagryansky, V. V. Prikhodko.} Formation
of a magnetic hole above the mirror-instability threshold in a
plasma with sloshing ions. Physical Review E. {\bf 81}, 067402
(2010). https://doi.org/10.1103/PhysRevE.81.067402

\bibitem{Skovorodin13}
{\it D. I. Skovorodin, K.V. Zaytsev, A. D. Beklemishev.} Global
sound modes in mirror traps with anisotropic pressure. Physics of
Plasmas {\bf 20}, 102123 (2013).

\bibitem{Casper82}
{\it T. A. Casper and G. R. Smith.} Observation of Alfv{\'e}n ion
cyclotron fluctuations in the end-cell plasma in the Tandem Mirror
Experiment. Phys. Rev. Lett. {\bf 48}, 15, 1015 (1982),
https://doi.org/10.1103/PhysRevLett.48.1015

\end{thebibliography}
\end{document}